\documentclass[11pt]{article}

\usepackage{amssymb}
\usepackage[numbers,sort&compress]{natbib}
\usepackage{amsmath,amsfonts,graphicx,epsfig}
\usepackage{bm}
\usepackage{color}

\def\bea{\begin{eqnarray}}
\def\eea{\end{eqnarray}}
\def\be{\begin{equation}}
\def\ee{\end{equation}}
\def\ba{\begin{array}}
\def\ea{\end{array}}

\def\nn{\nonumber}

\newcommand{\red}[1]{{\color{black} #1}}
\newcommand{\RED}[1]{{\color{black} #1}}

\begin{document}

\setlength\arraycolsep{2pt}

\renewcommand{\theequation}{\arabic{section}.\arabic{equation}}
\setcounter{page}{1}

\begin{titlepage}

\begin{center}

\vskip 1.5 cm

{\huge \bf Vanishing of local non-Gaussianity in canonical single field inflation}

\vskip 2.0cm

{\Large 
Rafael Bravo$^{a}$, Sander Mooij$^{a,b}$, Gonzalo A. Palma$^{a}$\\ and Basti\'an Pradenas$^{a}$
}

\vskip 0.5cm

{\it $^{a}$Grupo de Cosmolog\'ia y Astrof\'isica Te\'orica, Departamento de F\'{i}sica, FCFM, \mbox{Universidad de Chile}, Blanco Encalada 2008, Santiago, Chile. \\
$^{b}$Institute of Physics, Laboratory of Particle Physics and Cosmology, \'Ecole Polytechnique F\'ed\'erale de Lausanne, CH-1015 Lausanne, Switzerland  }

\vskip 2.5cm

\end{center}

\begin{abstract} 

We study the production of observable primordial local non-Gaussianity in two opposite regimes of canonical single field inflation: attractor (standard single field slow-roll inflation) and non attractor (ultra slow-roll inflation). In the attractor regime, the standard derivation of the bispectrum's squeezed limit using co-moving coordinates gives the well known \mbox{Maldacena's} consistency relation $f_{\rm NL} = 5 (1-n_s) / 12$. On the other hand, in the non-attractor regime, the squeezed limit offers a substantial violation of this relation given by $f_{\rm NL} = 5/2$. In this work we argue that, independently of whether inflation is attractor or non-attractor, the size of the \emph{observable} primordial local non-Gaussianity is predicted to be $f_{\rm NL}^{\rm obs} = 0$ (a result that was already understood to hold in the case of attractor models). To show this, we follow the use of the so-called Conformal Fermi Coordinates (CFC), recently introduced in the literature. These coordinates parametrize the local environment of inertial observers in a perturbed FRW spacetime, allowing one to identify and compute gauge invariant quantities, such as $n$-point correlation functions. Concretely, we find that during inflation, after all the modes have exited the horizon, the squeezed limit of the $3$-point correlation function of curvature perturbations vanishes in the CFC frame, regardless of the inflationary regime. We argue that such a cancellation should persist after inflation ends.

\end{abstract}

\end{titlepage}

\newpage

\setcounter{equation}{0}
\section{Introduction}

Maldacena's consistency relation~\cite{Maldacena:2002vr} has stood out as one of the key relations allowing us to test cosmic inflation~\cite{Guth:1980zm, Linde:1981mu, Starobinsky:1980te, Albrecht:1982wi, Mukhanov:1981xt}. It ties together two observables, the size of primordial local non-Gaussianity, $f_{\rm NL}$, and the power spectrum's spectral index, $n_s - 1$, in a simple relation given by:
\be
f_{\rm NL} = \frac{5}{12} (1 - n_s) . \label{s1-consistency-relation}
\ee
This relation has been shown to remain valid for all single-field attractor models of inflation, characterized by the freezing of the curvature perturbation after horizon crossing (attractor models of single field inflation are models in which every background quantity during inflation is determined by a single parameter, for instance, the value of the Hubble expansion rate $H$, regardless of the initial conditions). Moreover, eq.~(\ref{s1-consistency-relation}) is understood to be the consequence of how long wavelength modes of curvature perturbations modulate the amplitude of shorter wavelength modes~\cite{Creminelli:2004yq, Seery:2005wm, Chen:2006nt, Cheung:2007sv, Ganc:2010ff, RenauxPetel:2010ty,Kundu:2014gxa,Kundu:2015xta,Gong:2017yih}. This modulation is found to be enforced by a symmetry of the action for curvature perturbations under a transformation simultaneously involving spatial dilations and a field reparametrization.

Relation~(\ref{s1-consistency-relation}) is not satisfied by non-attractor models of single field inflation, for which the background depends crucially on the initial conditions~\cite{Tsamis:2003px, Kinney:2005vj, Namjoo:2012aa, Martin:2012pe, Mooij:2015yka}. For instance, in the extreme case of ultra slow-roll inflation~\cite{Tsamis:2003px, Kinney:2005vj}, where the amplitude of comoving curvature perturbations grows exponentially fast outside the horizon,\footnote{While the comoving curvature perturbation grows as $a^3$ on superhorizon scales, the non-adiabatic pressure still vanishes, providing a counterexample for the standard intuition that superhorizon freezing is caused by adiabaticity \cite{Romano:2015vxz}. See also \cite{Pajer:2017hmb}.} it has been shown~\cite{Namjoo:2012aa, Martin:2012pe, Mooij:2015yka, Bravo:2017wyw, Finelli:2017fml} that  $f_{\rm NL}$ is given by:\footnote{Ultra slow-roll is not a realistic model in at least two ways: (1) The inflationary potential is exactly flat, so inflation does not have an end. (2) The value of the spectral index is essentially zero. In fact, eq.~(\ref{s1-ultra-slow-roll}) may be written as $f_{\rm NL} = 5/2 - (1 - n_s) 5/4$, but  $n_s -1$ decays as $e^{- 6 N}$, where $N$ is the number of $e$-folds after horizon crossing. For these reasons, we take ultra slow-roll as a proxy model allowing for large local non-Gaussianity.}
\be
f_{\rm NL} = \frac{5}{2} . \label{s1-ultra-slow-roll}
\ee
This result has been extended to more general models of non-attractor inflation~\cite{Chen:2013aj}, for instance, realized with the help of nontrivial kinetic terms~\cite{Romano:2016gop}. Given that local non-Gaussianity is currently constrained by Planck as $f_{\rm NL} = 2.5 \pm 5.7$ (68\%CL)~\cite{Ade:2015ava}, the result shown in eq.~(\ref{s1-ultra-slow-roll}) has strengthened the importance of reaching new observational targets as a way to rule out (or to confirm) exotic mechanisms underlying the origin of primordial perturbations. Other models known for predicting potentially large local non-Gaussianity include multi-field models of inflation~\cite{Byrnes:2008wi} and curvaton models~\cite{Sasaki:2006kq}, but these scenarios require more than one scalar degree of freedom, so we will not consider them here.

Now, neither eq.~(\ref{s1-consistency-relation}) nor~(\ref{s1-ultra-slow-roll}) correspond to proper predictions for the amount of local non-Gaussianity available to inertial observers, such as us. In a cosmological setup, a physical local observer follows a geodesic path that does not necessarily coincide with that of an observer fixed at a given comoving coordinate. This is simply because inertial observers are themselves subject to the presence of perturbations. In the particular case of attractor models, when this fact is taken into account, one finds that the amount of local non-Gaussianity measured by an inertial observer is given by eq.~(\ref{s1-consistency-relation}) plus a correction of the same order. More to the point, one finds~\cite{Tanaka:2011aj, Pajer:2013ana, Dai:2015rda, Cabass:2016cgp, Tada:2016pmk}:
\bea
f^{\rm obs}_{\rm NL} =  0 . \label{s1-consistency-relation-observable}
\eea
In other words, the true prediction for observable primordial local non-Gaussianity coming from attractor models of inflation is zero.\footnote{More precisely, we should write $f^{\rm obs}_{\rm NL} =  0 + \mathcal O (k_L / k_S)^2$, where the $\mathcal O (k_L / k_S)^2$ terms are caused by non-primordial phenomena such as gravitational lensing and redshift perturbations (the so called projection effects~\cite{Baldauf:2011bh, Yoo:2009au}). See ref.~\cite{Pajer:2013ana} for more details on this.} Because of this, one may wonder about the status of the observable local non-Gaussianity in more general situations in which the value for local non-Gaussianity computed in comoving gauge is predicted to be large, just like in non-attractor models of inflation such as ultra slow-roll.

The purpose of this work is to analyze the prediction of observable local non-Gaussianity in more general contexts, beyond those covered by attractor models of single field inflation. To achieve this, we will adopt the Conformal Fermi Coordinates (CFC) formalism introduced by Pajer, Schmidt and Zaldarriaga in ref.~\cite{Pajer:2013ana} and further developed in refs.~\cite{Dai:2015rda, Cabass:2016cgp}. These coordinates are the natural generalization of the so-called Fermi normal coordinates~\cite{Manasse:1963zz}. In short, the use of CFC in a Friedmann-Robertson-Walker (FRW) spacetime allows one to describe the local environment of freely falling inertial observers up to distances much longer than the Hubble radius. This allows one to follow the fate of primordial curvature perturbations within the CFC frame during the whole relevant period of inflation, and perform the computation of gauge invariant $n$-point correlation functions. As we shall see, the expressions for $n$-point correlation functions in the CFC frame differ from those computed in comoving gauge. To understand why this is so, one has to keep in mind that while an ordinary gauge transformation (diffeomorphism) does not change physical observables, the introduction of CFC's corresponds to a change of coordinates that relate global with local coordinates.

We shall see that the long wavelength perturbations do not only modulate the evolution of perturbations of shorter wavelengths, they also affect the geodesic path of inertial observers. These two effects combined imply that long wavelength modes cannot affect short wavelength modes in any observable way. This has been well understood in the case of attractor models. Here we extend this result to the entire family of canonical single field inflation, regardless of whether they are attractor or non-attractor. To show this, we will compute the squeezed limit of the bispectrum in CFC coordinates, and find that it is given by
\be
B_{\rm CFC} = B_{\rm CM} + \Delta B, 
\ee
where $B_{\rm CM}$ is the bispectrum computed in comoving gauge, and $\Delta B$ is the correction due to the change of coordinates. The term $B_{\rm CM}$ is dictated by the invariance of the action for curvature perturbations under a transformation involving both spatial and time dilations together with a field reparametrization~\cite{Bravo:2017wyw}. On the other hand, we will show that the transformation dictating the form of $\Delta B$ is exactly the inverse of such a transformation. As a result, we find that during inflation, after all the modes have exited the horizon, $B_{\rm CFC}$ vanishes, regardless of whether inflation is attractor or non-attractor.

We begin this work by reviewing the construction of conformal Fermi coordinates in a perturbed FRW spacetime in Section~\ref{s2:CFC}. There we also sketch the computation of the squeezed limit of the primordial non-Gaussian bispectrum leading to $f^{\rm obs}_{\rm NL}$. Then, in Section~\ref{s3:FNL} we further develop and simplify some results found in refs.~\cite{Dai:2015rda, Cabass:2016cgp}. In addition, we apply these results to show that the observable primordial local non-Gaussianity vanishes in both, attractor and non-attractor models of inflation. Finally, in Section~\ref{s4:Conclusions} we offer some concluding remarks.

\setcounter{equation}{0}
\section{Review of Conformal Fermi Coordinates}
\label{s2:CFC}

Here we offer a review on how Conformal Fermi Coordinates (CFC) are defined and used to compute correlation functions for inertial observers in terms of those valid for comoving observers. We will mostly base this discussion on refs.~\cite{Dai:2015rda, Cabass:2016cgp}, with some slight variations of the notation. 

\subsection{Central geodesic}

Let us start by considering an unperturbed cosmological background described by a Friedmann-Robertson-Walker (FRW) metric in terms of conformal time $\tau$:
\be
ds_0^2 = a^2 (\tau) \eta_{\mu \nu} dx^\mu dx^\nu = a^2(\tau)  \left( - d \tau^2 + d {\bf x}^2 \right) .
\ee
Here $a(\tau)$ is the scale factor and ${\bf x}$ is the position in comoving coordinates. The perturbed spacetime may be described with the help of the following metric
\be
ds^2 = g_{\mu \nu} dx^\mu dx^\nu =  a^2 (\tau) \big[  \eta_{\mu \nu} + h_{\mu \nu} \big] dx^\mu dx^\nu  , \label{s2-perturbed-metric}
\ee
where $h_{\mu \nu}$ parametrize deviations from the FRW background. We will later use a specific form of $h_{\mu \nu}$ in which curvature perturbations are introduced. An inertial observer will follow a geodesic motion determined by $g_{\mu \nu}$, respecting the following equation of motion
\be
\frac{d^2 x^{\mu}}{d\eta^2} + \Gamma^{\mu}_{\rho \sigma} \frac{d  x^{\rho}}{d\eta} \frac{d  x^{\sigma}}{d\eta} = 0 ,
\ee
where $ \Gamma^{\mu}_{\rho \sigma}  = \frac{1}{2} g^{\mu \nu} (\partial_{\rho} g_{\nu \sigma}+ \partial_{\sigma} g_{\rho \nu} - \partial_{\nu} g_{\rho \sigma})$ are the usual Christoffel symbols, and $\eta$ is a given affine parameter. Let us call the resulting geodesic $G$, and take $\bar t$ to be the proper time employed by the inertial observer to parametrize time in his/her local environment. We will introduce a scale factor $a_F(\bar t)$ that parametrizes the expansion felt by the observer in his/her vicinity. The precise definition of $a_F(\bar t)$ is made in eq.~(\ref{s2-choice-a_F}). In the meantime, notice that the introduction of $a_F(\bar t)$ allows us to define a conformal time $\bar \tau$ through the following standard relation:
\be
d \bar \tau = d\bar t / a_F(\bar t) .
\ee
Because the inertial observer follows a geodesic motion that does not remain fixed to a comoving coordinate, it should be clear that $\tau$ and $\bar \tau$ will not coincide. Next, let us consider an arbitrary point $P$ along $G$ corresponding to conformal time $\bar \tau_P$. We wish to introduce a set of coordinates $x^{\bar \alpha} =  \left\{ \bar \tau , x^{\bar \imath} \right\}$ in such a way that $x^{\bar \imath}$ parametrizes the 3-dimensional slices of constant $\bar \tau_P$. In these coordinates, one has $x^{\bar \alpha} (P) = \left\{ \bar \tau_P , 0 \right\}$. At this point, one may introduce a set of tetrads $e^\mu_{\bar \alpha}$ such that:
\be
g_{\mu \nu} e^\mu_{\bar \alpha} e^\nu_{\bar \beta}   = \eta_{\bar \alpha \bar \beta} , \label{s2-tetrad-def}
\ee
in the vicinity of the entire geodesic. Equation~(\ref{s2-tetrad-def}) will be particularly true at the point $P$. We demand that the $\bar 0$-component $U^{\mu} \equiv  e^\mu_{\bar 0}$ coincides with the normalized vector tangent to the geodesic. Then, the rest of the tetrads $e^\mu_{\bar \imath}$ correspond to the space-like vectors orthogonal to the geodesic. 

In a perturbed FRW spacetime parametrized by the metric (\ref{s2-perturbed-metric}), the tetrads $e^\mu_{\bar \alpha}$ may be conveniently written as:
\bea
e^\mu_{\bar 0} &=&  \frac{1}{a (\tau)} \left( 1 + \frac{1}{2} h_{00} \, , \,V^{i} \right)  , \label{tetrad-pert-1} \\
 e^\mu_{\bar \jmath} &=&   \frac{1}{a (\tau)} \left( V_{j}  + h_{0 j} \, , \, \delta^{i}_{j} - \frac{1}{2} h^{i}{}_{j}  + \frac{1}{2} \varepsilon_{j}{}^{ik} \omega_k  \right) \delta^j_{\bar \jmath} ,\label{tetrad-pert-2}
\eea
where $V^{i}$ parametrizes the 3-component velocity of $U^{\mu}$, and $\omega_k$ parametrizes the rotation of the spatial components of the tetrad induced by the perturbations. In these expressions, and in the rest of this article, spatial indices are raised and lowered with $\delta^{ij}$ and $\delta_{ij}$ respectively. Notice that the construction outlined in the previous section requires that the combination $e^\mu_{\bar \alpha}$ be parallel transported along the geodesic (recall that $U^{\mu} = e^\mu_{\bar 0}$ and that $e^\mu_{\bar \imath}$ are normalized vectors orthogonal to $U^{\mu}$). Then $V^i$ and $\omega_k$ satisfy the following equations
\bea
\partial_0 V^i + \mathcal H V^i  &=& \frac{1}{2} \partial^i h_{00} - \partial_0 h^{i}{}_{0} - \mathcal H h^{i}{}_{0} ,  \label{s2-velocity-equation} \\
\partial_0 \omega^k &=& - \frac{1}{2} \varepsilon^{k i j} \left( \partial_i h_{0 j}  - \partial_j h_{0 i} \right) .  \label{s2-rotation-equation} 
\eea
In Section~\ref{CFC-inflation} we will see that $h_{0 i} = h_{i 0}$ is given by a gradient of the curvature perturbation [see eq.~(\ref{constraint-h-2})]. This will imply that the right hand side of eq.~(\ref{s2-rotation-equation}) vanishes, and $\omega^k$ may be chosen to vanish without loss of generality.

\subsection{Construction of the CFC map} \label{sec:Construction-CFC}

Now we have the challenge to define the slice of constant $\bar \tau_P$. An arbitrary point $Q$ in the vicinity of $P$, in the same slice of constant $\bar \tau_P$, will have coordinates $x^{\bar \alpha} (Q) = \{ \bar \tau_P , x^{\bar \imath}_Q \}$. We may reach $Q$ from $P$ through certain classes of geodesics that are constructed as follows: First, we introduce the conformally flat metric $\tilde g_{\mu \nu} \equiv a_F^{-2} (\bar \tau) g_{\mu \nu}$, and then solve the geodesic equation
\be
\frac{d^2 x^{\mu}}{d\lambda^2} + \tilde \Gamma^{\mu}_{\rho \sigma} \frac{d  x^{\rho}}{d\lambda} \frac{d  x^{\sigma}}{d\lambda} = 0 , \label{s2-geodesic-conformal}
\ee
where $ \tilde \Gamma^{\mu}_{\rho \sigma}$ are Christoffel symbols computed out of $\tilde g_{\mu \nu}$. To solve this equation, one chooses the following initial conditions:
\be
\frac{d  x^{\mu}}{d\lambda} \bigg|_{\lambda = 0} =  a_F(\bar \tau_P)   e_{\bar \imath}^{\mu} \Delta x^{\bar \imath}_{Q} , \label{s2-initial-condition}
\ee
where $\Delta x_Q^{\bar \imath} = x_Q^{\bar \imath} - x_P^{\bar \imath}$, and $x_Q^{\bar \imath}$ is the position of $Q$ that is reached at $\lambda = 1$. One may solve eq.~(\ref{s2-geodesic-conformal}) perturbatively by writing
\be
x^{\mu} (\lambda) = \sum_{n=0}^{\infty} \alpha_n^{\mu} \lambda^n .
\ee
Since $\lambda = 0$ corresponds to the starting point $P$, one has $ \alpha_0^{\mu} = x^{\mu} (P)$. On the other hand, the initial condition (\ref{s2-initial-condition}) implies $\alpha_1^{\mu} = e_{\bar \imath}^{\mu} \Delta x^{\bar \imath}_{Q} $. It is then possible to show that the solution to eq.~(\ref{s2-geodesic-conformal}), up to cubic order, is given by
\bea
x^{\mu} (\lambda) &=&  x^{\mu} (P) + a_F (\bar \tau_P)  e_{\bar \imath}^{\mu} \Big|_P \Delta x^{\bar \imath}_{Q} \, \lambda - \frac{1}{2}  \tilde \Gamma_{\rho \sigma}^{\mu}  a_F^{2}(\bar \tau_P)  e_{\bar \imath}^{\rho} e_{\bar \jmath}^{\sigma} \Big|_P  \Delta x^{\bar \imath}_{Q}   \Delta x^{\bar \jmath}_{Q}  \, \lambda^2 \nn \\ 
&& -   \frac{1}{6} \left( \partial_{\nu} \tilde \Gamma^{\mu}_{\rho \sigma} - 2 \tilde \Gamma^{\mu}_{\alpha \rho} \tilde \Gamma^{\alpha}_{\sigma \nu} \right) a_F^{3}(\bar \tau_P)  e_{\bar \imath}^{\rho} e_{\bar \jmath}^{\sigma} e_{\bar k}^{\nu} \Big|_P  \Delta x^{\bar \imath}_{Q}   \Delta x^{\bar \jmath}_{Q}  \Delta x^{\bar k}_{Q} \, \lambda^3 + \cdots .
\eea
Evaluating this result at $\lambda = 1$ then gives us the position of an arbitrary point $Q$ with respect to the position of $P$, and so, one may just drop the labels $P$ and $Q$ to obtain the coordinate transformation between the sets of coordinates $x^{\mu}$ and $x^{\bar \alpha}$ up to cubic order:
\bea
\Delta x^{\mu}  &=& a_F (\bar \tau_P)  e_{\bar \imath}^{\mu} \Big|_P  \Delta x^{\bar \imath} - \frac{1}{2}  \tilde \Gamma_{\rho \sigma}^{\mu}  a_F^{2}(\bar \tau_P)  e_{\bar \imath}^{\rho} e_{\bar \jmath}^{\sigma} \Big|_P  \Delta x^{\bar \imath}   \Delta x^{\bar \jmath} \nn \\
&&  - \frac{1}{6} \left( \partial_{\nu} \tilde \Gamma^{\mu}_{\rho \sigma} - 2 \tilde \Gamma^{\mu}_{\alpha \rho} \tilde \Gamma^{\alpha}_{\sigma \nu} \right) a_F^{3}(\bar \tau_P)  e_{\bar \imath}^{\rho} e_{\bar \jmath}^{\sigma} e_{\bar k}^{\nu} \Big|_P  \Delta x^{\bar \imath}   \Delta x^{\bar \jmath}  \Delta x^{\bar k} + \cdots, \label{s2:CFC-general-map}
\eea
where $\Delta x^{\mu} = x^{\mu} (Q) - x^{\mu} (P)$.
From this change of coordinates, it is now possible to deduce the form of the metric in conformal Fermi coordinates: 
\be
g_{\bar \alpha \bar \beta} = \frac{\partial x^\mu}{\partial x^{\bar \alpha}} \frac{\partial x^\nu}{\partial x^{\bar \beta}} g_{\mu \nu} .  \label{standard-coordinate-trans}
\ee
From this expression, one explicitly finds:
\bea
g_{\bar 0 \bar 0} &=& a_F^2 (\bar \tau) \left[ - 1 - \left( \tilde R_{\bar 0 \bar k \bar 0 \bar l} \right)_{P} \, \Delta x^{\bar k} \Delta x^{\bar l}  + \mathcal O (\Delta \bar x^3) \right] , \label{CFC-metric-1} \\
g_{\bar 0 \bar \jmath} &=& a_F^2 (\bar \tau) \left[ - \frac{2}{3} \left( \tilde R_{\bar 0 \bar k \bar \jmath \bar l} \right)_{P} \, \Delta x^{\bar k} \Delta x^{\bar l} + \mathcal O (\Delta \bar x^3) \right] , \label{CFC-metric-2} \\
g_{\bar \imath \bar \jmath} &=& a_F^2 (\bar \tau)  \left[ \delta_{ij} - \frac{1}{3} \left( \tilde R_{\bar \imath \bar k \bar \jmath \bar l} \right)_{P} \, \Delta x^{\bar k} \Delta x^{\bar l} + \mathcal O (\Delta \bar x^3) \right] , \label{CFC-metric-3}
\eea
where $\tilde R_{\bar \alpha \bar \beta \bar \gamma \bar \delta} $ are the components of the Riemann tensor constructed from $\tilde g_{\mu \nu}$ and projected along the CFC directions with the help of the tetrad introduced earlier:
\be
\tilde R_{\bar \alpha \bar \beta \bar \gamma \bar \delta} = a_F^{4}(\bar \tau_P)  e^{\mu}_{\bar \alpha} e^{\nu}_{\bar \beta} e^{\rho}_{\bar \gamma} e^{\sigma}_{\bar \delta} \tilde R_{\mu \nu \rho \sigma} .
\ee
In the previous expressions $\mathcal O (\Delta \bar x^3)$ stands for terms of order $\big( \partial_{\bar \imath} \tilde R_{\bar \alpha \bar \jmath \bar \beta \bar k} \big)_{P} \, \Delta  x^{\bar \imath} \Delta x^{\bar \jmath} \Delta x^{\bar k}$. This is in fact one of the salient points of this construction: Higher order corrections to the metric $g_{\bar \alpha \bar \beta}$ are suppressed by both spatial derivatives of the Riemann tensor and powers of $x^{\bar \imath}$. We will see how this plays a role later on, when we examine the validity of the CFC to follow the evolution of perturbations during inflation.

\subsection{Choosing the conformal scale factor $a_F$}

Notice that up to now the scale factor $a_F$ has not been properly defined. This may be done by first recalling that $U^{\mu} \equiv   e^\mu_{\bar 0}$ defined earlier satisfies the geodesic equation $U^{\nu} \nabla_\nu U^{\mu} = 0$. In order to study how two nearby parallel geodesics diverge, one may introduce the velocity divergence parameter $\vartheta$ as:
\be
\vartheta \equiv  \nabla_{\mu} U^{\mu} .
\ee
We then demand that the scale factor $a_F$ satisfies the following equation:
\be
H_F \equiv \frac{1}{a_F}  \frac{d a_F}{d \bar t} = \frac{1}{3} \vartheta . \label{s2-choice-a_F}
\ee
One crucial reason behind this choice is that $\vartheta$ is a local observable, and so $H_F$ is the true Hubble parameter describing the local expansion of the patch surrounding geodesic. One may read another consequence coming out from this choice. The velocity divergence $\vartheta$ satisfies the Raychaudhuri equation:
\be
\frac{d \vartheta}{d \bar t} +  \frac{1}{3} \vartheta^2=  - \sigma_{\mu \nu} \sigma^{\mu \nu} + \omega_{\mu \nu} \omega^{\mu \nu} - R^{\mu}{}_{\rho \mu \sigma} U^{\rho} U^{\sigma} ,
\ee
where $\sigma_{\mu \nu}$ and $\omega_{\mu \nu}$ are the trace-free symmetric and antisymmetric contributions to $\nabla_{\mu} U_{\nu}$ respectively. It is possible to work out $R^{\mu}{}_{\rho \mu \sigma} U^{\rho} U^{\sigma} = R^{\bar \mu}{}_{\bar \rho \bar \mu \bar \sigma} U^{\bar \rho} U^{\bar \sigma}$ to show that the Raychaudhuri equation reduces to
\be
\frac{d \vartheta}{d \bar t} +  \frac{1}{3} \vartheta^2= 3 (  \dot H_F + H_F^2 ) - \sigma_{\bar \imath \bar \jmath} \sigma^{\bar \imath \bar \jmath} + \omega_{\bar \imath \bar \jmath} \omega^{\bar \imath \bar \jmath} - a_F^{-2} \tilde R^{\bar \imath}{}_{\bar 0 \bar \imath \bar 0} .
\ee
But since $a_F$ has been chosen to satisfy (\ref{s2-choice-a_F}), this equation further reduces to
\be
 \sigma_{\bar \imath \bar \jmath} \sigma^{\bar \imath \bar \jmath} - \omega_{\bar \imath \bar \jmath} \omega^{\bar \imath \bar \jmath} = - a_F^{-2} \tilde R^{\bar \imath}{}_{\bar 0 \bar \imath \bar 0} .
\ee
In a homogeneous background, both $\sigma$ and $\omega$ vanish. Thus, we see that the choice of eq.~(\ref{s2-choice-a_F}) implies that $\tilde R^{\bar \imath}{}_{\bar 0 \bar \imath \bar 0}$ is necessarily of second order in perturbations.

\subsection{CFC in inflation} \label{CFC-inflation}

Now that we have in our hands the notion of conformal Fermi coordinates, we may examine their form in the specific case of a perturbed FRW spacetime during inflation. First, it is convenient to consider the perturbed metric of eq.~(\ref{s2-perturbed-metric}) in comoving gauge, where the coordinates are such that the fluctuations of the fluid driving inflation vanish. In the case of single field canonical inflation, this gauge corresponds to the case in which the perturbations of the scalar field driving inflation satisfy $\delta \phi = 0$. In this gauge, it is customary to introduce the comoving curvature perturbation variable $\zeta$ through the relation:
\be
h_{ij} = \left[ e^{2 \zeta} - 1 \right] \delta_{ij} . \label{curvature-comoving-def}
\ee 
Then, Einstein's equations imply constraint equations for $h_{00}$ and $h_{0i} = h_{i0}$. To linear order, the solutions of these equations are found to be given by
\bea
h_{00} &=& -  \frac{2}{\mathcal H} \partial_0 \zeta , \label{constraint-h-1} \\
h_{0i} &=& - \partial_i \left[ \frac{\zeta}{\mathcal H} - \epsilon \partial^{-2} \partial_0 \zeta  \right] , \label{constraint-h-2}
\eea
where $\mathcal H \equiv \partial_0 \ln a$ and  $\partial_0 \equiv \partial / \partial \tau$. These solutions will change in non-canonical models of inflation. For instance, the fluid driving inflation could induce an effective sound speed $c_s$ parametrizing the speed at which curvature perturbations propagate~\cite{Cheung:2007sv}, as in the case of $P(X)$-models~\cite{Garriga:1999vw} or single field EFT's describing multi-field models with massive fields~\cite{Achucarro:2010da}. In these cases, $c_s$ would modify the second constraint equation (\ref{constraint-h-2}).

Now, we would like to integrate eqs.~(\ref{s2-velocity-equation}) and~(\ref{s2-choice-a_F}) taking into account the introduction of the curvature perturbation $\zeta$. This will allow us to extend the CFC map (\ref{s2:CFC-general-map}) at any time $\tau \geq \tau_P$. Let us start by considering the integration of eq.~(\ref{s2-velocity-equation}) along the geodesic path. To do so, let us introduce the following combination involving $V^i$ and $h_{0i}$:
\be
\mathcal F^i  = V^i +  h_{0}{}^{i} . \label{V-F}
\ee
Given that $h_{0i}$ is a gradient, it should be clear that eq.~(\ref{s2-velocity-equation}) implies that the non-trivial part of $V^i$ is also given by a gradient. We therefore write $\mathcal F_i = \partial_i \mathcal F$. Then, direct integration of eq.~(\ref{s2-velocity-equation}) gives
\be
\mathcal F(\tau,{\bf x}) = e^{- \int^{\tau}_{\tau_*} ds \mathcal H (s)} \left[ \frac{1}{\mathcal H_*} C_{F} (\tau_* , {\bf x}) + \frac{1}{2} \int^{\tau}_{\tau_*} \!\!\! ds \, e^{\int^s_{\tau_*} dw \mathcal H (w)} h_{00} (s , {\bf x}) \right] , \label{F-integral-solution}
\ee
where $C_{F}(\tau_* , {\bf x}_c)$ is an integration constant defined on the geodesic path that must be taken to be linear in the perturbations. Given that we are interested in gradients of $\mathcal F$, we must allow for the existence of $\partial_i C_{F}(\tau_* , {\bf x})$ and $\partial^2 C_{F}(\tau_* , {\bf x})$. This result, together with $h_{0 i}$ found in (\ref{constraint-h-2}) gives us back $V^i$. Let us now consider the integration of eq.~(\ref{s2-choice-a_F}). By using eq.~(\ref{tetrad-pert-1}) to write $U^{\mu}$ in terms of $V^i$ and $h_{00}$, eq.~(\ref{s2-choice-a_F}) gives a first order differential equation for the combination $a_F (\bar \tau)/ a(\tau)$ which, to leading order in the perturbations yields
\be
\frac{a_F (\bar \tau)}{a (\tau)} - 1 =  C_{a} (\tau_* , {\bf x}_c(\tau_*))  + \int^{\tau}_{\tau_*} d s \left[ \partial_0 \zeta (s , {\bf x}_c (s) ) + \frac{1}{3} \partial_i V^i  (s , {\bf x}_c (s) )  \right] , \label{a/a-integral-solution}
\ee
where ${\bf x}_c (s)$ denotes the path of the geodesic in comoving coordinates. In the previous expression $\tau_*$ corresponds to a given initial time, and $C_{a}(\tau_* , {\bf x}_c)$ denotes an integration constant which should be considered to be of linear order in the perturbations.

We are now in a condition to deduce the form of the conformal Fermi coordinates valid at times $\tau > \tau_P$. To do so, let us consider an arbitrary point $P_2$ located on the central geodesic $G$ at a given time $\tau > \tau_P$. It is clear that $x^{\mu} (P_2 )$ will differ from the value $x^{\mu}_0 (P_2)$ that would have been obtained in an unperturbed universe. The difference is accounted by a deviation $\rho^{\mu} (\tau )$ that is at least linear in the perturbations:
\be
x^{\mu} (P_2 ) = x^{\mu}_0 (P_2) + \rho^{\mu} (\tau ) . \label{geodesic-dev}
\ee
Having this in mind, we may express the CFC map using the following ansatz for an arbitrary time $\tau > \tau_P$:
\be
x^{\mu} (\bar \tau , \bar {\bf x}) = x^{\mu}_0 (\bar \tau , \bar {\bf x}) + \rho^{\mu} (\tau) + A_{\bar \imath}^{\mu} (\tau) \Delta x^{\bar \imath} + B_{\bar \imath \bar \jmath}^{\mu} (\tau) \Delta x^{\bar \imath} \Delta x^{\bar \jmath}  + C_{\bar \imath \bar \jmath \bar k}^{\mu} (\tau) \Delta x^{\bar \jmath}  \Delta x^{\bar \jmath} \Delta x^{\bar k} + \cdots, \label{CFC-map-inflat}
\ee
where $x^{\mu}_0 (\bar \tau , \bar {\bf x})$ is the unperturbed map, for which the comoving coordinates and conformal Fermi coordinates coincide. More precisely, $x^{\mu}_0 (\bar \tau , \bar {\bf x})$  is such that:
\be
x^{0}_0 (\bar \tau , \bar {\bf x}) = \bar \tau  , \qquad x^{i}_0 (\bar \tau , \bar {\bf x}) =   x^i_c +  \delta^{i}_{\bar \imath}\Delta x^{\bar \imath} . \label{xmuzero}
\ee
The coefficients $ \rho^{\mu} (\tau )$, $A_{\bar \imath}^{\mu} (\tau)$, $B_{\bar \imath \bar \jmath}^{\mu} (\tau)$ and $C_{\bar \imath \bar \jmath \bar k}^{\mu} (\tau)$ are all linear in the perturbations. In Appendix~\ref{app:map-coeff} these are shown to be given by the following expressions:
\bea
 \rho^{0} (\tau )  &=&  \int^{\tau}_{\tau_*} \!\!\! ds \left[ \frac{a_F (\bar \tau)}{a (\tau)} (s , {\bf x}_c) - 1 + \frac{1}{2} h_{00} (s , {\bf x}_c) \right] , \label{map-coefficient-int-1} \\
  \rho^{i} (\tau )  &=&   \int^{\tau}_{\tau_*} \!\!\! ds \, V^{i}(s , {\bf x}_c) , \label{map-coefficient-int-2} \\
 A_{\bar \imath}^{0} (\tau) &=&  \delta^i_{\bar \imath} F_i (\tau , {\bf x}_c) , \label{map-coefficient-int-3} \\
  A_{\bar \imath}^{i} (\tau) &=& \left[ \frac{a_F (\bar \tau)}{a (\tau)}  - 1 - \zeta (\tau , {\bf x}_c) \right]  \delta^i_{\bar \imath} , \label{map-coefficient-int-4} \\
  B_{\bar \imath \bar \jmath}^{\mu} (\tau) &=& - \frac{1}{2} \tilde \Gamma^{\mu}_{i j } (\tau , {\bf x}_c) \delta^{i}_{\bar \imath} \delta^{j}_{\bar \jmath}, \label{map-coefficient-int-5} \\
  C_{\bar \imath \bar \jmath \bar k}^{\mu} (\tau) &=&  - \frac{1}{6} \partial_{k} \tilde \Gamma^{\mu}_{i j } (\tau , {\bf x}_c) \delta^{i}_{\bar \imath} \delta^{j}_{\bar \jmath} \delta^{j}_{\bar k} . \label{map-coefficient-int-6}
\eea
The right hand sides of the previous expressions are all expanded up to first order in the perturbations  (recall that $a_F (\bar \tau) / a (\tau) (s , {\bf x}_c) - 1$ is a quantity of linear order in the perturbations). Along the geodesic one has $\Delta \bar {\bf x} = 0$ and the map reduces to $x^{\mu} (\bar \tau , \bar {\bf x}_c) = x^{\mu}_0 (\bar \tau , \bar {\bf x}_c) + \rho^{\mu} (\tau )$. This implies that:
\be
\tau = \bar \tau + \rho^{0} (\tau ) , \qquad   x^i = x_c^i + \rho^i .
\ee
Then $\rho^{0} (\tau ) = \tau - \bar \tau$ informs us how the perturbations shift the equal time slices parametrized by $\tau$ and $\bar \tau$. Similarly, $\rho^i$ parametrizes the spatial shift of the geodesic from the unperturbed position $x_c^i$ (that is, $x_c^i + \rho^i$ is the location of the geodesic at a time $\tau > \tau_*$).

\subsection{Computation of correlation functions with CFC's} \label{correlations-in-CFC}

Here we consider the task of computing correlation functions using the CFC map of eq.~(\ref{CFC-map-inflat}). We are particularly interested in the squeezed limit of the three point function $\langle \zeta \zeta \zeta \rangle$. In this subsection we will sketch the procedure, that will be implemented in more detail in the next section, after we have considered some further simplifications of the map (\ref{CFC-map-inflat}). The main idea is the following: We will split the curvature perturbation $\zeta$ into short and long wavelength contributions:
\be
\zeta = \zeta_{S} + \zeta_{L} .
\ee
Then, we will use $\zeta_{L}$ to find the perturbed spacetime determining the map~(\ref{CFC-map-inflat}) deduced in the previous section. In other words, in global coordinates we study a FRW metric perturbed by $\zeta_L$. The map  of eq.~(\ref{CFC-map-inflat}) then shows that $\tau=\bar{\tau}+\mathcal{O}(\zeta_L)$, $x^i=\bar{x}^i+\mathcal{O}(\zeta_L)$. We then use the inverse of this map to see how local quantities (such as $\bar{\zeta}_S(\bar{x})$ and its correlation functions) can be written in terms of well-known global quantities. This will allow us to derive an expression for the short  wavelength curvature perturbation $\bar \zeta_S$ (defined in the inertial frame) as a function of both, the short and long wavelength curvature perturbations $\zeta_S$ and $\zeta_L$ (defined in the comoving frame):
\be
\bar \zeta_S =  \zeta_S + F_{\rm S} (\zeta_S ,\zeta_L) . \label{Split-1}
\ee
Here the function $F_{\rm S} (\zeta_S ,\zeta_L)$ informs us about how the long wavelength mode $\zeta_{L}$ affects the behavior of $\bar \zeta_S$ due to the fact that this is a local quantity defined within the patch surrounding the geodesic. This function will be of the form (see next subsection)
\be
F_{\rm S} (\zeta_S ,\zeta_L)  = \sum_a f_a ( \zeta_L) g_a (\zeta_S) , \label{F_S-form}
\ee
where $f_a ( \zeta_L)$ and $g_a (\zeta_S)$ are linear functions of $\zeta_L$ and $\zeta_S$ respectively (that could include space-times derivatives acting on the perturbations). Of course, the long mode $\zeta_{L}$ does not globally affect itself, in the sense that the CFC map corresponds to a local small scale coordinate transformation that can only affect the short scale contribution $\bar \zeta_S$ inside the patch around the central geodesic. Therefore we effectively write:
\be
\bar \zeta_L = \zeta_L .  \label{Split-2}
\ee
Having (\ref{Split-1}) and (\ref{Split-2}) then allows us to compute the squeezed limit of the three point correlation function $\langle \bar \zeta \bar \zeta \bar \zeta \rangle$. To this effect, one first considers the computation of the two point correlation function $\langle \bar  \zeta_S (\bar x_1) \bar  \zeta_S (\bar x_2) \rangle$. Because $\bar \zeta_S$ is given by (\ref{Split-1}) with $F_S$ given by (\ref{F_S-form}), this two point correlation function may be expanded to linear order in $\zeta_L$ as:
\be
\langle \bar  \zeta_S (\bar x_1) \bar  \zeta_S (\bar x_2) \rangle = \langle  \zeta_S ( x_1)  \zeta_S ( x_2) \rangle + \sum_a f_{a} (\zeta_{L}) \left[ \langle  \zeta_S (\bar x_1) g_a (\zeta_S (\bar x_2)) \rangle +  \langle  g_a (\zeta_S (\bar x_1)) \zeta_S (\bar x_2)  \rangle \right] .
\ee
Then, by correlating this result with the long mode $\bar \zeta_L$ of eq.~(\ref{Split-2}), one obtains
\bea
\langle \bar \zeta_{L} (\bar x_3) \langle \bar  \zeta_S (\bar x_1) \bar  \zeta_S (\bar x_2) \rangle \rangle = \langle \zeta_{L} ( x_3) \langle  \zeta_S ( x_1)  \zeta_S ( x_2) \rangle \rangle \qquad\qquad \qquad \qquad \qquad \qquad \qquad \nn \\
 +  \sum_a \langle \zeta_{L} ( x_3) f_{a} (\zeta_{L})  \rangle  \left[ \langle  \zeta_S (\bar x_1) g_a (\zeta_S (\bar x_2)) \rangle +  \langle  g_a (\zeta_S (\bar x_1)) \zeta_S (\bar x_2)  \rangle \right] . \label{three-point-corrected}
\eea
The squeezed limit in momentum space of this $\langle \bar \zeta_{L} (\bar x_3) \langle \bar  \zeta_S (\bar x_1) \bar  \zeta_S (\bar x_2) \rangle \rangle$ directly gives $f_{\rm NL}^{\rm obs}$, whereas the squeezed limit of the first term of the right hand side gives the usual $f_{\rm NL}$-parameter computed in comoving coordinates.\footnote{To show that $\langle \zeta_{L} ( x_3) \langle   \zeta_S ( x_1)  \zeta_S ( x_2) \rangle \rangle$ gives the squeezed limit of the bispectrum, one may start from $\langle \zeta ( x_3)   \zeta ( x_1)   \zeta ( x_2) \rangle$ and split both the curvature perturbation and the vacuum state as $\zeta = \zeta_L + \zeta_S$ and  $| 0 \rangle = | 0 \rangle_L \otimes | 0 \rangle_S$ respectively. Then one only needs to recall that, because of non-linearities, $\zeta_S$ will in ge\-ne\-ral depend on $\zeta_L$.} This means that one finally arrives to an expression of the form
\be
f_{\rm NL}^{\rm obs} = f_{\rm NL} + \Delta f_{\rm NL} ,
\ee
where $\Delta f_{\rm NL}$ arises from those terms entering the second line of eq.~(\ref{three-point-corrected}), due to the CFC transformation. We will see how to perform all these steps in detail in the following section.

\subsection{Short wavelength modes in CFC} \label{sec:splitting}

To finish this section, we deduce how long wavelength modes affect short wavelength modes through the CFC transformation of eq.~(\ref{CFC-map-inflat}). We will only consider the effects of the first three terms in~(\ref{CFC-map-inflat}), involving the perturbations $\rho^{\mu} (\tau )$ and $A_{\bar \imath}^{\mu} (\tau)$. The remaining pieces involving $B_{\bar \imath \bar \jmath}^{\mu} (\tau)$ and $C_{\bar \imath \bar \jmath \bar k}^{\mu} (\tau)$ introduce the so called projection effects, which are suppressed in the squeezed limit. We may organize the transformation as follows:
\be
x^{\mu} (\tau , \bar {\bf x}) = x^{\mu}_0 + \xi^{\mu} (\tau , \bar {\bf x})  , \label{CFC-map-inflat-needed}
\ee
where
\be
 \xi^{\mu} (\tau , \bar {\bf x}) \equiv \rho^{\mu} (\tau ) + A_{\bar \imath}^{\mu} (\tau) \Delta x^{\bar \imath} . \label{def-xi-tau-x}
\ee
Now, notice that the proper definition of the curvature perturbation in the CFC frame may be written as~\cite{Rigopoulos:2003ak, Lyth:2004gb}:
\be
\bar \zeta (\bar x) = \frac{1}{6} \log  \det (g_{\bar \imath \bar \jmath} / a_F^2 (\bar \tau) ) ,  \label{def-zeta-bar}
\ee
where the elements $g_{\bar \imath \bar \jmath}$ are given by the spatial components of eq.~(\ref{standard-coordinate-trans}) as:
\be
g_{\bar \imath \bar \jmath} = a^2 (\tau)\left[ \frac{\partial \tau}{\partial x^{\bar \imath}} \frac{\partial \tau}{\partial x^{\bar \jmath}} (1 + h_{00})
+  \frac{\partial x^i}{\partial x^{\bar \imath}} \frac{\partial \tau}{\partial x^{\bar \jmath}} h_{i 0}
+ \frac{\partial \tau}{\partial x^{\bar \imath}} \frac{\partial x^j}{\partial x^{\bar \jmath}} h_{0 j} 
+  \frac{\partial x^i}{\partial x^{\bar \imath}} \frac{\partial x^j}{\partial x^{\bar \jmath}} h_{ij} \right] . \label{standard-coordinate-trans-spatial}
\ee
We now consider the role of the long and short modes in the splitting $\zeta = \zeta_{L} + \zeta_{S}$ in order to compute $\bar \zeta (\bar x) $ from (\ref{def-zeta-bar}). First, notice that in the previous expression, the partial derivatives $\partial \tau / \partial x^{\bar \imath}$ and $\partial x^i / \partial x^{\bar \imath}$ are determined by the CFC map of eq.~(\ref{CFC-map-inflat}), and therefore depend on $\zeta_L$ alone. On the other hand, $h_{00}$, $h_{0i} = h_{i0}$ and $h_{ij} = \delta_{ij} e^{2 \zeta}$ depend on both $\zeta_{L}$ and $\zeta_{S}$. In addition, one has to keep in mind that $\zeta_S$ is evaluated at $x = x(\bar x)$, which is determined by the CFC map. Putting together all of these factors, it is possible to deduce
\be
\bar  \zeta_S (\bar x) = \zeta_S (x(\bar x))  + \frac{1}{3}  [h_{S}]_{0}{}^i (x(\bar x))  \partial_i \xi_L^0 (\bar x) ,  \label{zeta-s} 
\ee
where  the arguments of the short scale perturbations are evaluated at $x(\bar x)$. For example, in the case of the first term $\zeta_S(x(\bar x))$, since $x = x_0 + \xi$ (as in eq.~(\ref{CFC-map-inflat-needed})), one has
\be
\zeta_S (x(\bar x)) = \zeta_S (x_0) + \xi^\mu (\bar x) \partial_\mu \zeta_S (x_0) .
\ee
Recall that $x_0^{\mu}$ is given by eq.~(\ref{xmuzero}), and therefore $\zeta (x_0)$ is nothing but the comoving curvature perturbation evaluated with unperturbed comoving coordinates. For this reason, we could simply write $\zeta_S (x_0) = \zeta_S (\bar x)$.

\setcounter{equation}{0}
\section{Local non-Gaussianity in single field inflation}
\label{s3:FNL}

We now put together the results of the previous sections to compute the squeezed limit of the bispectrum in the two regimes of interest: attractor and non-attractor. To simplify matters, we will work to leading order in the slow-roll parameter $\epsilon$ and neglect any corrections that would modify the final results by terms suppressed in $\epsilon$.

\subsection{Further developments} \label{further-devs}

Let us start by obtaining explicit expressions for the integrals of eqs.~(\ref{F-integral-solution}) and~(\ref{a/a-integral-solution}), which in turn lead to simple and manageable expressions for the coefficients $\rho^{\mu} (\tau )$ and $A_{\bar \imath}^{\mu} (\tau)$ appearing in the map (\ref{CFC-map-inflat-needed}). First, notice that $\int^{\tau}_{\tau_*} \!\!\! ds \, \mathcal H (s) $ appearing in~(\ref{F-integral-solution}) may be directly integrated as:
\be
\int^{\tau}_{\tau_*} \!\!\! ds \, \mathcal H (s) =  \ln  \frac{a (\tau)}{a (\tau_*)} .
\ee
Then, one may re-express the integral of eq.~(\ref{F-integral-solution}) as
\be
 {\mathcal F} =   \frac{a(\tau_*)}{a(\tau)}  \left( \frac{1}{\mathcal H_*} C_F -  \left[  \frac{1}{a(\tau_*)}  \int^{\tau}_{\tau_*} \!\!\! d \tau \frac{a(\tau)}{\mathcal H (\tau)} \,   \frac{\partial \zeta}{\partial \tau} \right]  \right) . \label{F-N-int}
\ee
Now, notice that $\mathcal H = - 1 / \tau + \mathcal O (\epsilon)$ and $a (\tau) / a(\tau_*) =  \tau_* / \tau + \mathcal O (\epsilon)$. This allows one to integrate eq.~(\ref{F-N-int}) to obtain the following expression for $V_i$:
\bea
V_{i} &=&\frac{1}{\mathcal H_*} \frac{a(\tau_*)}{a(\tau)} \partial_i \Big( C_F +   \zeta_*  \Big) - \partial_i \left[ \epsilon \partial^{-2} \frac{\partial \zeta}{\partial \tau} \right] +  \mathcal O (\epsilon) , \label{F-integral-solution-3}
\eea
where $\mathcal O (\epsilon)$ stands for a function of order $\epsilon$ that decays quickly on superhorizon scales.  We will soon argue how to choose the integration constant $C_F$. Our choice will imply that $V_{i} $ is a function of order $\epsilon$. Irrespective of this, $V_{i}$ will contribute terms that quickly decay on superhorizon scales (for both regimes, attractor and non-attractor), and that become negligible in the computation of the bispectrum squeezed limit.  Next, we move on to compute the integral of eq.~(\ref{a/a-integral-solution}). Given that $V^i$ is sub-leading, eq.~(\ref{a/a-integral-solution}) may be directly integrated, giving us back

\be
\frac{a_F (\bar \tau)}{a (\tau)} -1 =   C_{a} + \zeta - \zeta_*  + \mathcal O (\epsilon) , \label{a/a-integral-solution-2}
\ee
where $ \mathcal O (\epsilon)$ stands for those decaying terms of order $\epsilon$. Finally, we may use these results to rewrite the map coefficients (\ref{map-coefficient-int-1})-(\ref{map-coefficient-int-4}) that will be used to compute the squeezed limit. Using the fact that $\mathcal H = - 1/ \tau$ up to corrections of order $\epsilon$,  these are found to be:
\bea
 \rho^{0} (\tau )  &=&    C_{a}  (\tau - \tau_*) +  \tau  ( \zeta - \zeta_*) + \cdots , \label{map-coefficient-pre-1}  \\
  \rho^{i} (\tau )  &=&  0 + \cdots , \label{map-coefficient-pre-2} \\
 A_{\bar \imath}^{0} (\tau) &=&  \delta^i_{\bar \imath}  \tau \partial_i \left[  \zeta - \zeta_*  - C_{F} \right] + \cdots, \label{map-coefficient-pre-3} \\
  A_{\bar \imath}^{i} (\tau) &=& \left[  C_{a}  - \zeta_*  \right]  \delta^i_{\bar \imath} + \cdots, \label{map-coefficient-pre-4}
\eea
where the ellipses $\cdots$ denote those decaying terms of order $\epsilon$.

\subsection{Initial conditions} \label{sec:initial-conditions}

As discussed in Section~\ref{correlations-in-CFC}, we are interested in understanding how long wavelength modes affect the geodesic motion of an inertial observer that has access to short wavelength perturbations. This means that the perturbed FRW spacetime considered in the previous section deviates from its unperturbed version due to long wavelength modes $\zeta_L$. We will choose $\tau_*$ at a time when all the relevant modes of $\zeta_L$ have exited the horizon. In practice, we are interested in computing the effects due to a single mode (or a small range of modes) appearing in $\zeta_L$, that will later be selected in the squeezed limit in momentum space. So we could simply say that our condition is that $\tau_*$ corresponds to a moment in time at which $\zeta_L$ has just crossed the horizon.

Given that at $\tau_*$ the perturbation $\zeta_L$ has just crossed the horizon, deviations to the geodesic path are just starting to take over. In particular, any effect of $\zeta_L$ on the velocity field $V_i$ must be negligible. We therefore choose $C_F$ in such a way that $V^i_* = 0$. To leading order this corresponds to 
\be
C_F = - \zeta^*_L . 
\ee
As explained, this implies that $V_i$ may be neglected, leading to $\rho^{i} (\tau ) = 0$, which was already stated in eq.~(\ref{map-coefficient-pre-2}). Next, a similar argument may be used to state that since $\zeta_L$ has just crossed the horizon, we require that $a (\tau_*) = a_F (\bar \tau_*)$, this corresponds to a synchronized map choice. This leads to $C_{a} = 0$ as evident from eq.~(\ref{a/a-integral-solution-2}). Then, we finally arrive at a simple version of the map coefficients needed to connect the coordinates of inertial and comoving observers, that may be summarized as follows:
\bea
 \rho^{0} (\tau )  &=&  \tau  ( \zeta_L - \zeta^{*}_L) + \cdots ,\label{map-coefficient-1}  \\
  \rho^{i} (\tau )  &=&  0 + \cdots , \label{map-coefficient-2} \\
 A_{\bar \imath}^{0} (\tau) &=&   \tau \partial_{\bar \imath}   \zeta_L  + \cdots, \label{map-coefficient-3} \\
  A_{\bar \imath}^{i} (\tau) &=&  - \zeta^*_L  \delta^i_{\bar \imath} + \cdots. \label{map-coefficient-4}
\eea
\RED{From eq.~(\ref{map-coefficient-1}) it is clear that in attractor inflation (where $\zeta_L$ freezes) time remains synchronized, while in the non-attractor case (where $\zeta_L$ evolves) the watches of a comoving and a free-falling observer do not run at the same rate.}

We notice that the authors of ref.~\cite{Cabass:2016cgp} discuss alternative choices to fix $\tau_*$ and the associated initial conditions.

\subsection{Computation of the squeezed limit}

We are finally ready to compute the observed bispectrum's squeezed limit. This computation was already outlined in Section~\ref{correlations-in-CFC}. We start by explicitly computing the two point correlation function $\langle \bar  \zeta_S (\bar x_1) \bar  \zeta_S (\bar x_2) \rangle$ using (\ref{zeta-s}) to express $\bar \zeta_S$ in terms of $ \zeta_S$:
\be
\langle \bar  \zeta_S (\bar x_1) \bar  \zeta_S (\bar x_2) \rangle = \left\langle \left(  \zeta_S (x(\bar x_1))  + \frac{1}{3} [h_{S}]_{0}{}^i \partial_i \xi_L^0  \right) \left(  \zeta_S (x(\bar x_2))  + \frac{1}{3} [h_{S}]_{0}{}^i \partial_i \xi_L^0  \right)  \right\rangle .
\ee
Notice that we have kept $x(\bar x)$ in the argument of $\zeta_S$ at the right hand side, which also depends on $\zeta_L$. We shall deal with this dependence in a moment. Expanding the previous expression up to linear order in $\xi^0_L$, we have
\bea
\langle \bar  \zeta_S (\bar x_1) \bar  \zeta_S (\bar x_2) \rangle &=& \langle \zeta_S (x(\bar x_1)) \zeta_S (x(\bar x_2)) \rangle  \nn \\�
&&+ \frac{1}{3} \left[ \left\langle  \zeta_S (x( \bar x_1))  [h_{S}]_{0}{}^i (\bar x_2)  \right\rangle  + \left\langle [h_{S}]_{0}{}^i  (\bar x_1)  \zeta_S (x(\bar x_2))   \right\rangle \right] \partial_i \xi_L^0  .
\eea
It is not difficult to show that, because $[h_{S}]_{0i}$ consists of a gradient, the two last terms of this expression cancel each other. Then, we are left with:
\be
\langle \bar  \zeta_S (\bar x_1) \bar  \zeta_S (\bar x_2)) \rangle = \left\langle \zeta_S (x(\bar x_1))     \zeta_S (x(\bar x_2))   \right\rangle .
\ee
Next, we may expand $x(\bar x)$ appearing in the argument of $\zeta_S$ in terms of $\zeta_L$. This gives:
\be
\langle \bar  \zeta_S (\bar x_1) \bar  \zeta_S (\bar x_2) \rangle = \left[ 1 + \xi_L^{\mu} (\tau , \bar {\bf x}_1) \partial^{(1)}_{\mu} + \xi_L^{\mu} (\tau , \bar {\bf x}_2) \partial^{(2)}_{\mu} \right] \left\langle  \zeta_S ( \bar x_1 )   \zeta_S ( \bar x_2 )   \right\rangle ,  \label{two-point-00}
\ee
where $\xi^{\mu} (\tau , \bar {\bf x})$ is given in eq.~(\ref{def-xi-tau-x}), and where $\partial^{(1)}_{\mu}$ and $\partial^{(2)}_{\mu}$ are partial derivatives with respect to $\bar x_1$ and $\bar x_2$ respectively. As already explained in Section~\ref{correlations-in-CFC}, the two point correlation function $\left\langle  \zeta_S ( \bar x_1 )   \zeta_S ( \bar x_2 )   \right\rangle$ depends on $\zeta_L$. But given that $\xi^{\mu}$ in (\ref{two-point-00}) already depends linearly on $\zeta_L$, in order to keep the leading terms, we may re-write the previous expression as
\be
\langle \bar  \zeta_S (\bar x_1) \bar  \zeta_S (\bar x_2) \rangle = \left\langle  \zeta_S ( \bar x_1 )   \zeta_S ( \bar x_2 )   \right\rangle  + \left[ \xi_L^{\mu} (\tau , \bar {\bf x}_1) \partial^{(1)}_{\mu} + \xi_L^{\mu} (\tau , \bar {\bf x}_2) \partial^{(2)}_{\mu} \right] \left\langle  \zeta_S ( \bar x_1 )   \zeta_S ( \bar x_2 )   \right\rangle_0 ,  \label{two-point-0}
\ee
where $\left\langle  \zeta_S ( \bar x_1 )   \zeta_S ( \bar x_2 )   \right\rangle_0$ is the two point correlation function in comoving coordinates with $\zeta_L \to 0$ (that is, without a modulation coming from the long wavelength mode). Notice that $\left\langle  \zeta_S ( \bar x_1 )   \zeta_S ( \bar x_2 )   \right\rangle_0$  is nothing but the two point correlation function of $\zeta_S$ with spatial arguments given by $\bar x_1$ and $\bar x_2$ as if it was computed in comoving coordinates. The result is a function of time $\tau$, and the difference $|\bar {\bf x}_1- \bar {\bf x}_2|$:
\be
\left\langle  \zeta_S ( \bar x_1 )   \zeta_S ( \bar x_2 )   \right\rangle_0 = \left\langle  \zeta_S    \zeta_S  \right\rangle (\bar{\tau} , \bar{r}) , \qquad r \equiv |\bar {\bf x}_2- \bar {\bf x}_1| . \label{spatial-homogeneity}
\ee
Using this result back into eq.~(\ref{two-point-0}) together with the map coefficients of eqs.~(\ref{map-coefficient-1})-(\ref{map-coefficient-1}), we find (see Appendix~\ref{app:two-point} for details):
\bea
\langle \bar  \zeta_S (\bar x_1) \bar  \zeta_S (\bar x_2) \rangle  &=&  \left\langle  \zeta_S ( \bar x_1 )   \zeta_S ( \bar x_2 )   \right\rangle \nn \\ 
&& + \left[ ( \zeta_L - \zeta^{*}_L)  \frac{\partial}{\partial \ln \tau} + \frac{1}{2}  ( x_1^{\bar \imath} + x_2^{\bar \imath}) \partial_{\bar \imath}  \zeta_L  \frac{\partial}{\partial \ln \tau}     - \zeta^*_L   \frac{\partial}{\partial \ln r}  \right] \left\langle  \zeta_S    \zeta_S  \right\rangle (\tau , r)  . \qquad    \label{two-point-S}
\eea
Recall that $\zeta_L$ is evaluated at ${\bf x}_c$. However, given that it is a long wavelength mode, we may as well consider it to be evaluated at $\bar {\bf x}_L = (\bar {\bf x}_1 + \bar {\bf x}_2)/2$ without modifying any conclusion. Note that the second term inside the square brackets is necessarily subleading since it involves a spatial derivative of the long-wavelength mode $\zeta_L$. For this reason, we disregard it. To continue, we may now Fourier transform this expression. First, we introduce
\be
\zeta ({\bf x}) = \frac{1}{(2 \pi)^3} \int d^3 k \zeta({\bf k}) e^{ i {\bf k} \cdot {\bf x}} ,
\ee
which implies that
\bea
 \langle \bar  \zeta_S (\bar x_1) \bar  \zeta_S (\bar x_2) \rangle = \left\langle  \zeta_S ( \bar x_1 )   \zeta_S ( \bar x_2 )   \right\rangle  \qquad \qquad \qquad \qquad \qquad \qquad \qquad \qquad \qquad \qquad \qquad    \nn \\ 
+ \int \!\! \frac{d^3 k}{(2 \pi)^3}e^{ i {\bf k} \cdot {\bf x}_L} \left( \left[ \zeta({\bf k})  -  \zeta_*({\bf k})  \right]    \frac{\partial}{\partial \ln \tau}   -   \zeta_*({\bf k})   \frac{\partial}{\partial \ln r}  \right) \left\langle  \zeta_S    \zeta_S  \right\rangle (\tau , r).   \label{zeta-L-SS-Fourier}
\eea
Then, Fourier transforming the fields $\bar  \zeta_S (\bar x_1)$ and $\bar  \zeta_S (\bar x_2)$, we arrive to (see Appendix~\ref{app:two-point} for details)
\bea
 \langle    \bar \zeta_S \bar \zeta_S \rangle ({\bf k}_1 ,{\bf k}_2)  =  \langle \zeta_S \zeta_S \rangle ({\bf k}_1 ,{\bf k}_2)  \qquad   \qquad \qquad \qquad \qquad \qquad \qquad \qquad \qquad  \quad  \nn \\ 
+     \left( \left[ \zeta({\bf k}_L)  -  \zeta_*({\bf k}_L)   \right]    \frac{\partial}{\partial \ln \tau}   +   \zeta_*({\bf k}_L)  \left[ n_s(k_S , \tau)- 1 \right] \right)  P_\zeta (\tau, k_S) ,  \label{CFC-effect-power}
\eea
where ${\bf k}_L = {\bf k}_1 + {\bf k}_2$ and ${\bf k}_S = ( {\bf k}_1 - {\bf k}_2 ) / 2$. In the previous expressions, the power spectrum $P_\zeta (\tau, k) $ of $\zeta ({\bf k})$ and its spectral index $n_s (k) - 1$ are defined as
\bea
P_\zeta (\tau, k)  =  \int d^3 r  e^{ - i  {\bf k} \cdot  {\bf r}} \left\langle  \zeta    \zeta  \right\rangle (\tau , r) , \\
n_s(k , \tau)- 1 =  \frac{\partial }{\partial \ln k} \ln ( k^3 P_\zeta (\tau, k) ) .
\eea
Equation (\ref{CFC-effect-power}) gives the power spectrum in conformal Fermi coordinates expressed in terms of the curvature perturbations defined in comoving coordinates. 

To continue, notice that since we have split the curvature perturbation in short and long wavelength modes, the two point correlation function $\langle \zeta_S \zeta_S \rangle ({\bf k}_1 ,{\bf k}_2) $ will be modulated by the long wavelength mode $\zeta_L$ in comoving coordinates. The squeezed limit of the bispectrum in comoving coordinate appears as the formal limit:
\be
\lim_{k_3 \to 0}(2 \pi)^3 \delta ({\bf k}_1 + {\bf k}_2 +{\bf k}_3) B_{\zeta}  ({\bf k}_1 , {\bf k}_2 , {\bf k}_3)  = \langle  \zeta_L ({\bf k}_3) \langle \zeta_S \zeta_S \rangle ({\bf k}_1 ,{\bf k}_2)  \rangle .
\ee
Thus, we see that if we correlate the expression of eq.~(\ref{CFC-effect-power}) with $\zeta_L ({\bf k}_3)$ we obtain (after using eq.~(\ref{Split-2}))
\bea
\langle  \bar \zeta_L ({\bf k}_3)  \langle    \bar \zeta_S \bar \zeta_S \rangle ({\bf k}_1 ,{\bf k}_2) \rangle  &=& (2 \pi)^3 \delta ({\bf k}_1 + {\bf k}_2 +{\bf k}_3) B_{\zeta}   ({\bf k}_1 , {\bf k}_2 , {\bf k}_3)   \nn \\ 
&& +   (2 \pi)^3 \delta ({\bf k}_1 + {\bf k}_2 +{\bf k}_3) P_\zeta (\tau, k_3) \frac{\partial}{\partial \ln \tau} P_\zeta (\tau, k_S) \nn \\  
&&  -   \langle \zeta_L ({\bf k}_3)  \zeta_*({\bf k}_L) \rangle \left[  \frac{\partial}{\partial \ln \tau} -   \left[ n_s(k_S , \tau)- 1 \right] \right] P_\zeta (\tau, k_S) . \label{3-point-1}
\eea
\RED{Here we still work with the notation  ${\bf k}_L = {\bf k}_1 + {\bf k}_2$ and ${\bf k}_S = ( {\bf k}_1 - {\bf k}_2 ) / 2$.} Notice that this expression contains the quantity $ \langle   \zeta_L ({\bf k}_3)  \zeta_*({\bf k}_L) \rangle $, which correlates two $ \zeta_L $ at two different times. In what follows, we show that 
\be
\langle \bar \zeta_L ({\bf k}_3)  \langle    \bar \zeta_S \bar \zeta_S \rangle ({\bf k}_1 ,{\bf k}_2) \rangle = 0 , \label{3-point-2}
\ee
for both attractor and non-attractor models of inflation.

\subsection{Vanishing of local non-Gaussianity} \label{sec:vanishing_of_NG}

In a recent article~\cite{Bravo:2017wyw} we have derived a generalized version of the non-Gaussian consistency relation valid for the two regimes of interest: attractor and non-attractor models. The squeezed limit of the 3-point correlation function for $\zeta$ was found to be given by:\footnote{In~\cite{Finelli:2017fml} a different expression for the generalization of the squeezed limit was obtained. The derivation of~\cite{Finelli:2017fml} is based on the use of the operator product expansion to find the squeezed limit of the three point functions for $\zeta$. The main difference with the result (\ref{power-S-corr-2}) found in~\cite{Bravo:2017wyw} consists of the presence of terms with two $\ln \tau$-derivatives, instead of one.}
\bea
\langle  \zeta_L ({\bf k}_3)  \langle  \zeta_S \zeta_S \rangle ({\bf k}_1 ,{\bf k}_2) \rangle &=&  - \langle  \zeta_L ({\bf k}_3)  \left[  \zeta_L ({\bf k}_L) - \zeta_L^* ({\bf k}_L )\right] \rangle  \frac{d}{d \ln \tau}   P_\zeta (\tau, k_S) \nn \\
&& - \langle  \zeta_L ({\bf k}_3)  \zeta_L^* ({\bf k}_L)  \rangle \left[ n_s(k_S , \tau)- 1 \right]  P_\zeta (\tau, k_S) ,  \label{power-S-corr-2}
\eea
where the fields $\zeta_L$ and $\zeta_S$ are evaluated at a time $\tau$, and $\zeta_L^*$ is evaluated at a reference initial time $\tau_*$. In order to derive this expression, we used the fact that the cubic action for the curvature perturbation $\zeta$ is approximately invariant under space-time reparametrizations given by 
\bea
&& x \to x'  = e^{  \zeta_L (\tau_*) }  x , \label{rep-4} \\ 
&& \tau \to \tau' = e^{- \zeta_L (\tau) + \zeta_L (\tau_*)  } \tau  . \label{rep-5} 
\eea
It may be seen how these relations resemble the coordinate transformation implied by eqs.~(\ref{map-coefficient-1})-(\ref{map-coefficient-4}), except for the signs of $\zeta_L (\tau)$ and $\zeta_L (\tau_*)$.

In the case of attractor models, one has that the comoving curvature perturbation $\zeta$ becomes constant on superhorizon scales. This implies that $\zeta_*({\bf k}_L) = \zeta_L ({\bf k}_3) $ and that $\ln \tau$-derivatives of the power spectrum vanish. Then eq.~(\ref{power-S-corr-2}) reduces to
\bea
B_{\zeta}   ({\bf k}_1 , {\bf k}_2 , {\bf k}_3)  &=&  -  \left[ n_s(k_S , \tau)- 1 \right]  P_\zeta (\tau, k_L) P_\zeta (\tau, k_S) .
\eea
This is the well known Maldacena's consistency condition. On the other hand, in non-attractor models of inflation (ultra slow-roll) the modes grow exponentially fast on superhorizon scales, and one finds a leading contribution given by (one finds that $n_s(k_3 , \tau)- 1 \propto \epsilon_* (\tau / \tau_*)^6$ and so it may be regarded as formally zero in the long wavelength limit): 
\bea
B_{\zeta}   ({\bf k}_1 , {\bf k}_2 , {\bf k}_3)  &=&  -  P_\zeta (\tau, k_L) \frac{d}{d \ln \tau} P_\zeta (\tau, k_S) .   \label{power-S-corr-2-leading}
\eea
More precisely, the superhorizon modes evolve like $\zeta ({\bf k}, \tau) =  (\tau_* / \tau)^3 \zeta ({\bf k}, \tau_*) $~\cite{Kinney:2005vj, Namjoo:2012aa, Martin:2012pe, Mooij:2015yka}. For this reason, the power spectrum scales as $\tau^{-6}$:
\be
P_\zeta (\tau, k) = \left(\frac{\tau_*}{ \tau} \right)^6 P_\zeta (\tau_*, k) .
\ee
In this case, eq.~(\ref{power-S-corr-2-leading}) finally gives
\be
B_{\zeta}   ({\bf k}_1 , {\bf k}_2 , {\bf k}_3) = 6 P_\zeta (\tau , k_L ) P_\zeta (\tau, k_S),
\ee 
which is the well known ultra slow-roll bispectrum~\cite{Namjoo:2012aa}.

Now, independently of the specific form of the bispectrum in these two regimes, we see that eq.~(\ref{power-S-corr-2}) implies a cancellation between $B_{\zeta} ({\bf k}_1 , {\bf k}_2 , {\bf k}_3)$ and the rest of the terms in eq.~(\ref{3-point-1}):
\be
B_{\zeta}   ({\bf k}_1 , {\bf k}_2 , {\bf k}_3) + \Delta B_{\zeta}   ({\bf k}_1 , {\bf k}_2 , {\bf k}_3) = 0
\ee
Thus, we conclude that in the CFC frame the bispectrum vanishes \red{during both the attractor and non-attractor regimes. In the next section we will argue why we expect this result to survive after inflation ends.}

\subsection{On the validity of CFC for non-attractor models}

Let us briefly come back to the issue raised in Section~\ref{sec:Construction-CFC}, regarding the validity of the CFC map in the case of non-attractor models.  We note that even if $\zeta$ grows as $a^3$ on superhorizon scales, during inflation it is always small, since it should reach its value observed in the CMB $\zeta_{\rm CMB} \simeq 10^{-5}$ when it stops to evolve, i.e., at the end of inflation. 

Therefore, the validity of the CFC transformation does not rely on the size of $\zeta$ but on the possibility that the map depends on time derivatives of $\zeta$, which are of order $\mathcal H \zeta$. To be more precise, notice from eqs.~(\ref{CFC-metric-1})-(\ref{CFC-metric-3}) that the size of the patch surrounding the central geodesic G has to be such that 
\be
\left|  \left( \tilde R_{\bar 0 \bar k \bar 0 \bar l} \right)_{P} \, \Delta x^{\bar k} \Delta x^{\bar l} \right| \ll 1 ,  \label{eq:condition-1}
\ee
where $\tilde R_{\bar 0 \bar k \bar 0 \bar l}$ are components of the Riemann tensor constructed from the conformally flat metric $\tilde g_{\mu \nu} \equiv a_F^{-2} (\bar \tau) g_{\mu \nu}$. This means that $\tilde R_{\bar 0 \bar k \bar 0 \bar l}$ is at least linear in the perturbations. Then, eq.~(\ref{eq:condition-1}) simply translates into the condition that 
 $|\Delta x|^2 \mathcal H^2 \zeta_L \ll  1$, where we have used that time derivatives of $\zeta$ are of order $\mathcal{H}\zeta$.  This last inequality is guaranteed by our previous remark on the size of $\zeta$. 
 
Next, one could be worried about higher order time derivatives of $\zeta$ emerging in terms of order, for instance those of order $\mathcal O (\Delta \bar x^3)$ that appear in eqs.~(\ref{CFC-metric-1})-(\ref{CFC-metric-3}). However, as noted before, higher order contributions that are linear in $\tilde R_{\bar 0 \bar k \bar 0 \bar l}$ (and therefore linear in $\zeta$) only bring in spatial derivatives with respect to $x^{\bar \imath}$. Higher order derivatives with respect to time will only come about from terms which are quadratic in $\tilde R_{\bar 0 \bar k \bar 0 \bar l}$, and are thus quadratic in $\zeta_L^2$. Therefore, even if in ultra slow-roll $\zeta$ does grow exponentially on superhorizon scales, the convergence of both these expansions is still guaranteed.

\setcounter{equation}{0}
\section{Discussion and conclusions}
\label{s4:Conclusions}

We have studied the computation of local non-Gaussianity accessible to inertial observers in canonical models of single field inflation. It was already known~\cite{Tanaka:2011aj, Pajer:2013ana, Cabass:2016cgp, Tada:2016pmk} that observable local non-Gaussianity vanishes in the case of single field attractor models ($f_{NL}^{\rm obs} = 0$) modulo projection effects. In this work, we have extended this result to the case of non-attractor models (ultra slow-roll) in which the standard derivation gives a sizable value $f_{NL} = 5/2$. This result (the standard result) was thought to represent a gross violation of Maldacena's consistency relation. We have instead shown that for both classes of models, the consistency relation is simply:
\be
f_{\rm NL}^{\rm obs} = 0. \label{conc-main-result}
\ee 
This result is noteworthy: In ultra slow-roll comoving curvature perturbations experience an exponential superhorizon growth, and this growth was taken to be  the natural explanation underlying large local non-Gaussianity. But this is certainly not the case. 

Our results shed new light on our understanding of the role of the bispectrum squeezed limit in inflation to test primordial cosmology. We now know that non-Gaussianity cannot discriminate between two drastically different regimes of inflation. Instead, we are forced to think of new ways of testing the evolution of curvature perturbations in non-attractor backgrounds. This is particularly important once we face the possibility that ultra slow-roll could be representative of a momentary phase within a conventional slow-roll regime~\cite{Germani:2017bcs, Dimopoulos:2017ged}.

In order to derive (\ref{conc-main-result}), we have re-examined the use of conformal Fermi coordinates introduced in ref.~\cite{Pajer:2013ana} and perfected in refs.~\cite{Dai:2015rda, Cabass:2016cgp}. Our results complement these works. For instance, the vanishing of $f_{\rm NL}^{\rm obs}$ in the case of non-attractor models required us to consider in detail the contribution of time-displacements of the CFC map that are irrelevant in the case of attractor models. 

The previous remark offers a way to understand the vanishing of local $f_{\rm NL}^{\rm obs}$ for the case of non-attractor models. To appreciate this, let us first focus on the case of attractor models. Notice that in the case of attractor models the freezing of the curvature perturbation can be absorbed at superhorizon scales through a re-scaling of the \RED{spatial}  coordinates, which, to linear order in the perturbations, looks like ${\bf x} \to {\bf x}' = {\bf x} + \zeta_* {\bf x}$, where $\zeta_*$ is the value of the mode at horizon crossing. It is precisely this scaling that gives rise to the modulation of small scale perturbations by long scale perturbations in comoving gauge. The map coefficients of eq.~(\ref{map-coefficient-1}) show that in attractor inflation the local transformation corresponds to ${\bf x} \to \bar {\bf x} = {\bf x} - \zeta_* {\bf x}$. This transformation is opposite to the previous re-scaling, and therefore it cancels the effect of the modulation in comoving coordinates. \RED{Note that all these transformations act only on spatial coordinates, the time coordinate remains untouched in the attractor case.}

Now, something similar happens in the case of non-attractor models. Here, the curvature perturbation does not freeze on superhorizon scales. Instead, on superhorizon scales the mode acquires a time dependence that may be absorbed by a re-scaling of time $\tau \to \tau' = \tau  - \zeta(\tau) \tau$ in the argument of the scale factor (in comoving coordinates). Similar to the case of attractor models, the map coefficients of eq.~(\ref{map-coefficient-1}) show that in the non-attractor regime the local CFC transformation corresponds to $\tau \to \bar \tau = \tau + \zeta(\tau) \tau$, which is opposite to the previous re-scaling, and so it cancels the whole modulation effect.

More generally, and independently of whether we are looking into the attractor regime, or the non-attractor regime, the cancellation may be understood as follows: The squeezed limit of the 3-point correlation function of canonical models of inflation is the consequence of a symmetry of the action for $\zeta$ under the special class of space-time reparametrization shown in eqs.~(\ref{rep-4})-(\ref{rep-5}). This symmetry is exact in the two regimes that we have studied, but approximate in intermediate regimes. In addition, this symmetry dictates the way in which long-wavelength $\zeta$-modes modulate their short wavelength counterparts. The CFC transformation is exactly the inverse of the symmetry transformation, and so the modulation deduced with the help of the symmetry is cancelled by moving into the CFC frame. 

At this point, it is important to emphasize that our computation was performed during inflation. That is, we have performed the CFC transformation while inflation takes place, and the result $B + \Delta B = 0$ found in Section~\ref{sec:vanishing_of_NG} is strictly valid during inflation. The claim that the primordial contribution to $f_{\rm NL}^{\rm obs}$ vanishes for a late time observer must be a consequence of the CFC transformation, taking into account the entire cosmic history. This would require studying the transition from the non-attractor phase to the next phase, which presumably could be of the attractor class, a study already begun in \cite{Cai:2016ngx}. \red{Note that the non-attractor nature of USR inflation leads to many different ways to end this phase. However, given that in both regimes (pure ultra slow-roll and pure slow-roll inflation), we have seen that }both $B$ and $\Delta B$ found in Section~\ref{sec:vanishing_of_NG} are exactly the same (but of opposite signs) and determined by $\tau$-derivatives and $k$-derivatives of the power spectrum, we expect that the end of non-attractor inflation (which could be a transition to a slow-roll phase) will affect equally $B$ and $\Delta B$, in such a way that the net result \red{will} continue to be $B + \Delta B = 0$. Verifying this claim, which seems reasonable, is however out of the scope of the present article.\footnote{In a recent article~\cite{Cai:2017bxr} (written simultaneously to this work), Cai \emph{et al.} studied the effects on the bispectrum $B$ of a transition from a non-attractor phase to an attractor phase. They discovered that the transition can drastically change the \red{comoving} value of $f_{\rm NL}$, suppressing its value if the transition is smooth. Then, the question would be: what happens with $\Delta B$ during such transitions? \red{Does the transition from comoving to Fermi coordinates continue to cancel the squeezed bispectrum computed in comoving coordinates? As we explain above, we expect the answer to be affirmative.}}

\red{Our main reason for expecting that }local non-Gaussianity \red{when expressed in free-falling Fermi vanishes} in both attractor and non-attractor models of single field inflation \red{ is that} in both cases, after the inflaton scalar degree of freedom is swallowed by the gravitational field, the only dynamical scalar degree of freedom corresponds to the curvature perturbation. As a consequence, the interaction coupling together long and short wavelength modes is purely gravitational, and therefore the equivalence principle dictates that long wavelength physics cannot dictate the evolution of short wavelength dynamics, implying that any observable effect must be suppressed by a ratio of scales $\mathcal O (k_L / k_S)^2$. All of this calls for a better examination of the relation between the local ansatz and the squeezed limit of the bispectrum~\cite{dePutter:2016moa}.

Our work leaves several open challenges ahead. First, we have focussed our interest in ca\-no\-ni\-cal models of inflation, namely, those in which the inflaton field is parametrized by a Lagrangian containing a canonical kinetic term. In this category, the ultra slow-roll regime is not fully realistic, and at best should be considered as a toy model allowing the study of perturbations under the extreme conditions of a non-attractor background. However, it has been shown that \RED{non}-attractor regimes may appear more realistically within non-canonical models of inflation such as $P(X)$ models. In these models one has non-gravitational interactions inducing a sound speed $c_s \neq 1$, and so we suspect that our result (\ref{conc-main-result}) will not hold in those cases. \RED{This intuition is mainly based on the fact that in non-attractor models, the squeezed limit gets an enhancement when $c_s^2 \neq 1$ as shown in Ref.~\cite{Chen:2013aj}.  At any rate}, this work (together with ref.~\cite{Bravo:2017wyw}) calls for a better understanding of the non-Gaussianity predicted by non-attractor models in general.

Second, given that observable local non-Gaussianity vanishes in ultra slow-roll, in which curvature perturbations grow exponentially on superhorizon scales, one should revisit the status of other classes of inflation, such as multi-field inflation, where local non-Gaussianity may be large (a first look into this issue has already been undertaken in ref.~\cite{Tada:2016pmk}). It is quite feasible that in some models of multi-field inflation the amount of local non-Gaussianity may be understood as the consequence of a space-time symmetry dictating the way in which long-wavelength modes module short modes. 

Third, a deeper understanding of our present result is in order. In the case of attractor models, Maldacena's consistency relation (and its vanishing) may be understood as a consequence of soft limit identities linking the non-linear interaction of long wavelength perturbations with shorter ones~\cite{Creminelli:2012ed, Hinterbichler:2012nm, Senatore:2012wy, Assassi:2012zq, Hinterbichler:2013dpa, Goldberger:2013rsa, Creminelli:2013cga}. However, there were good reasons to suspect that these relations would not hold anymore in the case of non-attractor models~\cite{Pajer:2013ana}. Our results suggest that, regardless of the background, these identities continue to be valid, and in an inertial frame the gravitational interaction cannot be responsible of making long wavelength modes affect the local behavior of short wavelength modes. 

\subsection*{Acknowledgements}
We would like to thank Xingang Chen, Enrico Pajer, Dong-Gang Wang and Mat\'ias Zaldarriaga for helpful discussions and comments. GAP acknowledges support from the Fondecyt Regular project number 1171811. B.P. acknowledges the CONICYT-PFCHA Magister Nacional Scholarship 2016-22161360. RB also acknowledges the support from CONICYT-PCHA Doctorado Nacional  scholarship 2016-21161504. SM is funded by the Fondecyt 2015 Postdoctoral Grant 3150126.

\begin{appendix}

\setcounter{equation}{0}
\renewcommand{\theequation}{\Alph{section}.\arabic{equation}}
\section{Details of computations}

In this appendix we provide details of various intermediate steps performed in the bulk of the work. 

\subsection{Map coefficients} \label{app:map-coeff}

Here we show how to compute the map coefficients appearing in eqs.~(\ref{map-coefficient-int-1})-(\ref{map-coefficient-int-6}). To start with, we notice that the map of eq.~(\ref{CFC-map-inflat}) implies that along the geodesic (where $\Delta x^{\bar \imath} = 0$) the following relations must be satisfied:
\bea
\frac{\partial \tau}{\partial x^{\bar \imath}} &=& A^0_{\bar \imath} , \label{coef-der-1} \\
\frac{\partial x^i}{\partial x^{\bar \imath}} &=& A^i_{\bar \imath} , \label{coef-der-2} \\
\frac{\partial x^i}{\partial \bar \tau} &=& \frac{\partial \rho^{i}}{\partial \bar \tau} . \label{coef-der-3}
\eea
In addition, it is useful to recall the transformation rule of eq.~(\ref{standard-coordinate-trans}) connecting the metric in comoving coordinates and conformal Fermi coordinates:
\be
g_{\bar \alpha \bar \beta} = \frac{\partial x^\mu}{\partial x^{\bar \alpha}} \frac{\partial x^\nu}{\partial x^{\bar \beta}} g_{\mu \nu} .  \label{standard-coordinate-trans-app}
\ee
At any point on the geodesic we have that $g_{\bar \alpha \bar \beta} = a_{F}^2 (\bar \tau) \eta_{\bar \alpha \bar \beta}$. Then, given that the tetrads satisfy $g_{\mu \nu} e^\mu_{\bar \alpha} e^\nu_{\bar \beta}   = \eta_{\bar \alpha \bar \beta}$, we see that eqs.~(\ref{coef-der-1}) and~(\ref{coef-der-2}) imply 
\bea
A^0_{\bar \imath} &=& a_F(\bar \tau) e^{0}_{\bar \imath} , \\
A^i_{\bar \imath} &=& a_F(\bar \tau) e^{i}_{\bar \imath} .
\eea
Then, using the expression of eq.~(\ref{tetrad-pert-2}) for the tetrad $e^{\mu}_{\bar \imath}$, and keeping the leading contributions in terms of the perturbations, we find
\bea
A^0_{\bar \imath} &=&  \delta^i_{\bar \imath}  \partial_{i} \mathcal F, \label{A-0-i-app} \\
A^i_{\bar \imath} &=&  \left[ \frac{a_F (\bar \tau)}{a (\tau)} - 1 - \zeta  \right]  \delta^i_{\bar \imath}  ,
\eea
where we used the fact that $a_F (\bar \tau) / a (\tau) - 1$ is of linear order. These correspond to the desired expressions for the map coefficients (\ref{map-coefficient-int-3}) and (\ref{map-coefficient-int-4}).

Next, let us consider eq.~(\ref{standard-coordinate-trans-app}) with the choice $(\bar \alpha , \bar \beta) = (\bar 0 , \bar 0)$. Because along the geodesic one has $g_{\bar 0 \bar 0} = - a_F^2 (\bar \tau)$, using the form of the metric $g_{\mu \nu}$ introduced in eq.~(\ref{s2-perturbed-metric}) one arrives at
\be
\frac{a_F^2 (\bar \tau)}{a^2 (\tau)} = \left( \frac{\partial \tau}{\partial \bar \tau} \right)^2 \left[ 1 - h_{00} \right] - 2 \frac{\partial \tau }{\partial \bar \tau} \frac{\partial x^i}{\partial \bar \tau} h_{0 i}  - \frac{\partial x^i}{\partial \bar \tau} \frac{\partial x^j}{\partial \bar \tau} \delta_{ij} e^{2 \zeta} .
\ee
Keeping the leading terms, the previous expression may be rewritten as:
\be
 \frac{\partial \tau}{\partial \bar \tau}  = \frac{a_F (\bar \tau)}{a (\tau)}  + \frac{1}{2} h_{00}  .
\ee
Then, integrating with respect to time we arrive to:
\be
\tau - \tau_*  = \int^{\tau}_{\tau_*} ds \left[ \frac{a_F (\bar \tau)}{a (\tau)}  + \frac{1}{2} h_{00} \right]  . \label{app-pre-integration}
\ee
As a last step, note that $\rho^{0} = \tau - \bar \tau$ by definition. This implies that:
\be
\rho^0  = \int^{\tau}_{\tau_*} ds \left[ \frac{a_F (\bar \tau)}{a (\tau)} - 1  + \frac{1}{2} h_{00} \right]  .
\ee
which is the desired expression giving us the map coefficient of eq.~(\ref{map-coefficient-int-1}). Notice that we have imposed the condition $\rho^0 = 0$ at $\tau = \tau_*$ to synchronize the map. 

To conclude, let us consider eq.~(\ref{standard-coordinate-trans-app}) one more time for the case $(\bar \alpha , \bar \beta) = (\bar 0 , \bar \imath)$:
\be
0 =   \frac{\partial \tau}{\partial \bar \tau} \frac{\partial \tau}{\partial x^{\bar \imath}}  \left[  - 1 + h_{00} \right]  +  \frac{\partial x^i }{\partial \bar \tau} \frac{\partial \tau }{\partial x^{\bar \imath}} h_{i 0}+  \frac{\partial \tau }{\partial \bar \tau} \frac{\partial x^j}{\partial x^{\bar \imath}} h_{0 j}   + \frac{\partial x^i}{\partial \bar \tau} \frac{\partial x^j}{\partial x^{\bar \imath}} \delta_{ij} e^{2 \zeta} .
\ee
Keeping the leading terms in the perturbations, we obtain 
\be
\frac{\partial \rho^i}{\partial \bar \tau} = A^0_{\bar \imath} \delta^{i \bar \imath}  - \delta^{ij}  h_{0 j}   .
\ee
Then, inserting our previous result of eq.~(\ref{A-0-i-app}) and integrating with respect to time, we finally arrive to:
\be
\rho^{i}  =   \int^{\tau}_{\tau_*} \!\!\! ds \, V^{i} ,
\ee
which corresponds to the map coefficient of eq.~(\ref{map-coefficient-int-2}). 

Finally, the map coefficients of eqs.~(\ref{map-coefficient-int-5}) and (\ref{map-coefficient-int-6}) directly follow directly from eq.~(\ref{s2:CFC-general-map}) after keeping the leading therms in the perturbations.

\subsection{The 2-point correlation function for short modes}  \label{app:two-point}

Here we give details on how to derive eqs.~(\ref{two-point-S}) and (\ref{CFC-effect-power}). Let us start with eq.~(\ref{two-point-S}). First, notice that the second term at the right hand side of eq.~(\ref{two-point-0}) may be rewritten as
\bea
 \left[ \xi_L^{\mu} (\tau , \bar {\bf x}_1) \partial^{(1)}_{\mu} + \xi_L^{\mu} (\tau , \bar {\bf x}_2) \partial^{(2)}_{\mu} \right] \left\langle  \zeta_S ( \bar x_1 )   \zeta_S ( \bar x_2 )   \right\rangle_0 
  =    \bigg[ \frac{1}{2}(  \xi_L^{0} (\tau , \bar {\bf x}_2) - \xi_L^{0} (\tau , \bar {\bf x}_1)  )  ( \partial^{(2)}_{0} - \partial^{(1)}_{0} ) \quad \nn \\
  +  \frac{1}{2} \bigg( \xi_L^{0} (\tau , \bar {\bf x}_1)  +  \xi_L^{0} (\tau , \bar {\bf x}_2) \bigg) \partial_{0}  + \xi_L^{i} (\tau , \bar {\bf x}_1) \partial^{(1)}_{i} + \xi_L^{i} (\tau , \bar {\bf x}_2) \partial^{(2)}_{i}\bigg] \left\langle  \zeta_S ( \bar x_1 )   \zeta_S ( \bar x_2 )   \right\rangle_0 . \qquad \label{two-point-app-1}
\eea
Let us focus for a moment on the first contribution of the right hand side, which is given by 
\be
\frac{1}{2}A^0_{\bar \imath} ( x_2^{\bar \imath} - x_1^{\bar \imath} )   \left\langle ( \zeta_S ( \bar x_1 )   \zeta'_S ( \bar x_2 )  - \zeta'_S ( \bar x_1 )   \zeta_S ( \bar x_2 )  ) \right\rangle_0, \label{app:first-contrib}
\ee
where we have used $\xi_L^{0} (\tau , \bar {\bf x}_2) - \xi_L^{0} (\tau , \bar {\bf x}_1) = A^0_{\bar \imath} ( x_2^{\bar \imath} - x_1^{\bar \imath} )$. Now recall that the label $0$ reminds us that we are dealing with a comoving curvature perturbation in the absence of non-linearities. We may therefore expand it as 
\be
 \zeta_S ( \bar x ) = \frac{1}{(2 \pi)^3} \int dk e^{- i {\bf k} \cdot {\bf x}} \zeta (\tau , {\bf k}) , \qquad  \zeta (\tau , {\bf k}) = \zeta_{k} (\tau) a_{\bf k} + \zeta_{k}^* (\tau) a_{-\bf k}^{\dag} ,
\ee
where $\zeta_{k} (\tau)$ are is the amplitude of the mode, and $a_{\bf k}^{\dag}$ and $a_{\bf k}$ are creation and annihilation operators. I is then easy to show that 
\be
\left\langle  \zeta (\tau , {\bf k}_1) {\zeta^*}' (\tau , {\bf k}_2)  - \zeta' (\tau , {\bf k}_1)  \zeta^* (\tau , {\bf k}_2)   \right\rangle_0 = (2 \pi)^3 \delta^{(3)} ({\bf k}_2 - {\bf k}_1) [ \zeta_{k_1}' (\tau) \zeta_{k_1}^* (\tau) - \zeta_{k_1} (\tau)  {\zeta_{k_1}^*}' (\tau)] .
\ee
Now, $[ \zeta_{k_1}' (\tau) \zeta_{k_1}^* (\tau) - \zeta_{k_1} (\tau)  {\zeta_{k_1}^*}' (\tau)]$ must be such that the canonical commutation condition for $\zeta$ is satisfied. This implies that (\ref{app:first-contrib}) is given by
\be
\frac{1}{2}A^0_{\bar \imath} ( x_2^{\bar \imath} - x_1^{\bar \imath} )   \left\langle ( \zeta_S ( \bar x_1 )   \zeta'_S ( \bar x_2 )  - \zeta'_S ( \bar x_1 )   \zeta_S ( \bar x_2 )  ) \right\rangle_0 =\frac{1}{4 \epsilon a^2}A^0_{\bar \imath} ( x_2^{\bar \imath} - x_1^{\bar \imath} )  \delta^{(3)} ({\bf x_2} - {\bf x_1}),
\ee
which gives vanishing contribution to (\ref{two-point-app-1}). Next, using (\ref{spatial-homogeneity}) and replacing in $\xi_L^{\mu}$ the map coefficients of eqs.~(\ref{map-coefficient-int-1})-(\ref{map-coefficient-int-4}), we find that the remaining terms in eq.~(\ref{two-point-app-1}) give:
\bea
 &&\left[ \xi_L^{\mu} (\tau , \bar {\bf x}_1) \partial^{(1)}_{\mu} + \xi_L^{\mu} (\tau , \bar {\bf x}_2) \partial^{(2)}_{\mu} \right] \left\langle  \zeta_S ( \bar x_1 )   \zeta_S ( \bar x_2 )   \right\rangle_0 
  \nn \\�
 && \quad  =    \bigg[  \frac{1}{\tau} \left(  \rho^{0} +   \frac{1}{2}  ( x_1^i + x_2^i )A^{0}_{i} \right) \frac{\partial}{\partial \ln \tau}   + A^i_{\bar \imath} ( x_2^{\bar \imath} - x_1^{\bar \imath} ) (x_2^i - x_1^i) \frac{1}{r^2}  \frac{\partial}{\partial \ln r} \bigg]   \left\langle  \zeta_S    \zeta_S  \right\rangle ( \tau  , r)  . \qquad    \label{two-point-app-2}
\eea
This result then leads directly to eq.~(\ref{two-point-S}).

Next, we wish to derive eq.~(\ref{CFC-effect-power}) out from eq.~(\ref{zeta-L-SS-Fourier}). The first step is to simply multiply eq.~(\ref{zeta-L-SS-Fourier}) by $e^{ - i {\bf k}_1 \cdot \bar {\bf x}_1- i {\bf k}_2 \cdot \bar {\bf x}_2}$ and integrate the two spatial coordinates $\bar x_1$ and $\bar x_2$. This directly gives
\bea
 \langle    \bar \zeta_S \bar \zeta_S \rangle ({\bf k}_1 ,{\bf k}_2) &=& \langle \zeta_S \zeta_S \rangle ({\bf k}_1 ,{\bf k}_2) + \int d^{3} \bar x_1 d^{3} \bar x_2 e^{ - i {\bf k}_1 \cdot \bar {\bf x}_1- i {\bf k}_2 \cdot \bar {\bf x}_2} \int \!\! \frac{d^3 k}{(2 \pi)^3}e^{ i {\bf k} \cdot {\bf x}_L} \nn \\
&& \times \left( \left[ \zeta({\bf k})  -  \zeta_*({\bf k}) +  \zeta({\bf k}) i {\bf k} \cdot {\bf x}_L \right]    \frac{\partial}{\partial \ln \tau}   -   \zeta_*({\bf k})   \frac{\partial}{\partial \ln r}  \right) \left\langle  \zeta_S    \zeta_S  \right\rangle (\tau , r).  \qquad \label{app:zeta-L-SS-Fourier-1}
\eea
We may re-express the integral in terms of $\bf r = \bf {x}_1 - \bf {x}_2$ and ${\bf x}_L = ( \bf {x}_1+ \bf {x}_2) / 2$ instead of $\bar {\bf x}_1$ and $\bar {\bf x}_2$. One has that $d^{3} \bar x_1 d^{3} \bar x_2 = d^3 r d^3 x_L$, and so one finds
\bea
 \langle    \bar \zeta_S \bar \zeta_S \rangle ({\bf k}_1 ,{\bf k}_2) &=& \langle \zeta_S \zeta_S \rangle ({\bf k}_1 ,{\bf k}_2) + \int d^{3} r d^{3} x_L e^{ - i {\bf k}_S  \cdot {\bf r} / 2 } \int \!\! \frac{d^3 k}{(2 \pi)^3} e^{ i ( {\bf k} - {\bf k}_L) \cdot {\bf x}_L  } \nn \\
&& \times \left( \left[ \zeta({\bf k})  -  \zeta_*({\bf k}) +  \zeta({\bf k}) i {\bf k} \cdot {\bf x}_L \right]    \frac{\partial}{\partial \ln \tau}   -   \zeta_*({\bf k})   \frac{\partial}{\partial \ln r}  \right) \left\langle  \zeta_S    \zeta_S  \right\rangle (\tau , r),  \qquad \label{app:zeta-L-SS-Fourier-2}
\eea
where we have defined ${\bf k}_L = {\bf k}_1 + {\bf k}_2$ and ${\bf k}_S = ( {\bf k}_1 - {\bf k}_2 ) / 2$. The final step consists of explicitly integrating the coordinates ${\bf x}_L$ and $\bf r$. The $r$-integral gives us the power spectrum $P_\zeta (\tau, k_S)$. On the other hand, the $x_L$-integral gives us a $\delta^{(3)} ( {\bf k} - {\bf k}_L)$ that can be used to get rid of the $k$-integral (and produces the appearance of $\nabla_{{\bf k}_L}$). All of this together gives us the final result expressed in eq.~(\ref{CFC-effect-power}), after we throw away the term suppressed by ${\bf k}_L$.

\end{appendix}

\end{document}